\documentclass[prl,reprint,amssymb,amsmath,aps,superscriptaddress,showpacs,showkeys]{revtex4-1}
\usepackage{bm}
\usepackage{dcolumn}
\usepackage[linkcolor=blue]{hyperref}
\usepackage{graphicx}
\usepackage{color}
\usepackage{url}

\begin{document}

\preprint{SHOCK WAVES ON COMPLEX NETWORKS}

\begin{abstract}
Power grids, road maps, and river streams are examples of infrastructural networks which are highly vulnerable to external perturbations. An abrupt local change of load (voltage, traffic density, or water level) might propagate in a cascading way and affect a significant fraction of the network.
Almost discontinuous perturbations can be modeled by shock waves which can eventually interfere constructively and endanger the normal functionality of the infrastructure.
We study their dynamics by solving the Burgers equation under random perturbations on several real and artificial directed graphs.
Even for graphs with a narrow distribution of node properties (e.g., degree or betweenness), a steady state is reached exhibiting a heterogeneous load distribution, having a difference of one order of magnitude between the highest and average loads.
Unexpectedly we find for the European power grid and for finite Watts-Strogatz networks a broad pronounced bimodal distribution for the loads.
To identify the most vulnerable nodes, we introduce the concept of node-basin size, a purely topological property which we show to be strongly correlated to the average load of a node.
\end{abstract}


\title[Shock waves on complex networks]{ Shock waves on complex networks }
\author{ Enys Mones }
\email{ enys@hal.elte.hu }
\affiliation{ Department of Biological Physics, E\"{o}tv\"{o}s Lor\'{a}nd University,
P\'{a}zm\'{a}ny P\'{e}ter S\'{e}t\'{a}ny. 1/A, H-1117 Budapest, Hungary }
\affiliation{ Computational Physics for Engineering Materials, IfB, ETH Z\"{u}rich,
Wolfgang-Pauli-Strasse 27, CH-8093 Z\"{u}rich, Switzerland }
\date{\today}
\author{ Nuno A. M. Ara\'{u}jo }
\email{ nuno@ethz.ch }
\affiliation{ Computational Physics for Engineering Materials, IfB, ETH Z\"{u}rich,
Wolfgang-Pauli-Strasse 27, CH-8093 Z\"{u}rich, Switzerland }
\author{ Tam\'{a}s Vicsek }
\email{ vicsek@hal.elte.hu }
\affiliation{ Department of Biological Physics, E\"{o}tv\"{o}s Lor\'{a}nd University,
P\'{a}zm\'{a}ny P\'{e}ter S\'{e}t\'{a}ny. 1/A, H-1117 Budapest, Hungary }
\affiliation{ Biological Physics Research Group of HAS, P\'{a}zm\'{a}ny P\'{e}ter stny. 1/A, H-1117 Budapest, Hungary }
\author{ Hans J. Herrmann }
\email{ hans@ifb.baug.ethz.ch }
\affiliation{ Computational Physics for Engineering Materials, IfB, ETH Z\"{u}rich,
Wolfgang-Pauli-Strasse 27, CH-8093 Z\"{u}rich, Switzerland }
\affiliation{ Departamento de F\'{i}sica, Universidade Federal do Cear\'{a},
60451-970 Fortaleza, Cear\'{a}, Brazil }
\maketitle

\section{ Introduction }
\label{sec:introduction}
Blackouts, traffic gridlocks, and floods are all malfunctions of infrastructures which drastically affect their performance \cite{brummit12, helbing01, mamede12, lerner09, anand12}.
In many situations, they occur abruptly and might propagate through the network as shock waves \cite{zeldovich02,rudenberg68,calvert11,gai10}.
These waves can either weaken by shedding their impact among branches or interfere constructively when two or more branches meet at the same node.
The global consequences of these perturbations will strongly depend on the propagation dynamics and the capacity of each network element to bear abrupt changes \cite{moreira09,allen10,schneider11}.
The identification of vulnerable spots is a challenging scientific and technological question and this is precisely what we address here.

Propagation of failures and cascading in complex networks have been subject of much scientific interest \cite{watts02,sachtjen00,kosterev99}.
Examples are the use of the theory of self-organized criticality to study the propagation of failures in power grids and water transport on reservoir networks \cite{bak87,bonabeau95,goh03,brummit12,mamede12,araujo13,noel13}, the Olami--Feder--Christensen model for earthquakes \cite{olami92,lise02}, traffic \cite{chowdhury00,helbing01,mendes12} and financial networks \cite{battiston12}.
Typically, the focus is on the cascading of failures resulting from an initial triggering event.
However, it is also crucial to understand the dynamics preceeding these failures and identify the vulnerable spots where they can possibly be triggered.

To describe the propagation of shock waves on directed networks we use the Burgers equation \cite{burgers74}.
This equation describes flow when the flux depends quadratically on the load (e.g., voltage, traffic density, and water level).
The range of applications of the Burgers equation goes beyond fluid dynamics as it is applied in many propagation processes, such as traffic jams, glacier avalanches or chemical processes \cite{whitham74,tas07}.
Here we show that, in the case of perturbations randomly distributed in space, the dynamics of the solutions of the dissipative Burgers equation converges to a steady state in which the load distribution is strongly heterogeneous.
Surprisingly, we find that the load of some nodes can exceed the average load by one order of magnitude.
One might expect that the location of such nodes mainly depends on the propagation dynamics.
Yet, we show that their fate is deeply imprinted in the network topology.
We propose a new topological measure which allows to identify the most vulnerable nodes without solving the dynamics.

\section{ Model }
\label{sec:model}
\emph{Dynamics}---To describe the propagation of load (e.g., traffic density or water level) on a directed network, we consider on each link the one-dimensional Burgers equation \cite{whitham74}
\begin{equation}
    \frac{\partial\rho}{\partial t}+\rho\frac{\partial\rho}{\partial x}=0,
    \label{eq:1d_burgers_equation}
\end{equation}
which we solve using Godunov's scheme.
The details of the discretization and numerical solution are presented in the section Methods.
\begin{figure*}[t!]
  \includegraphics[width=17.2cm]{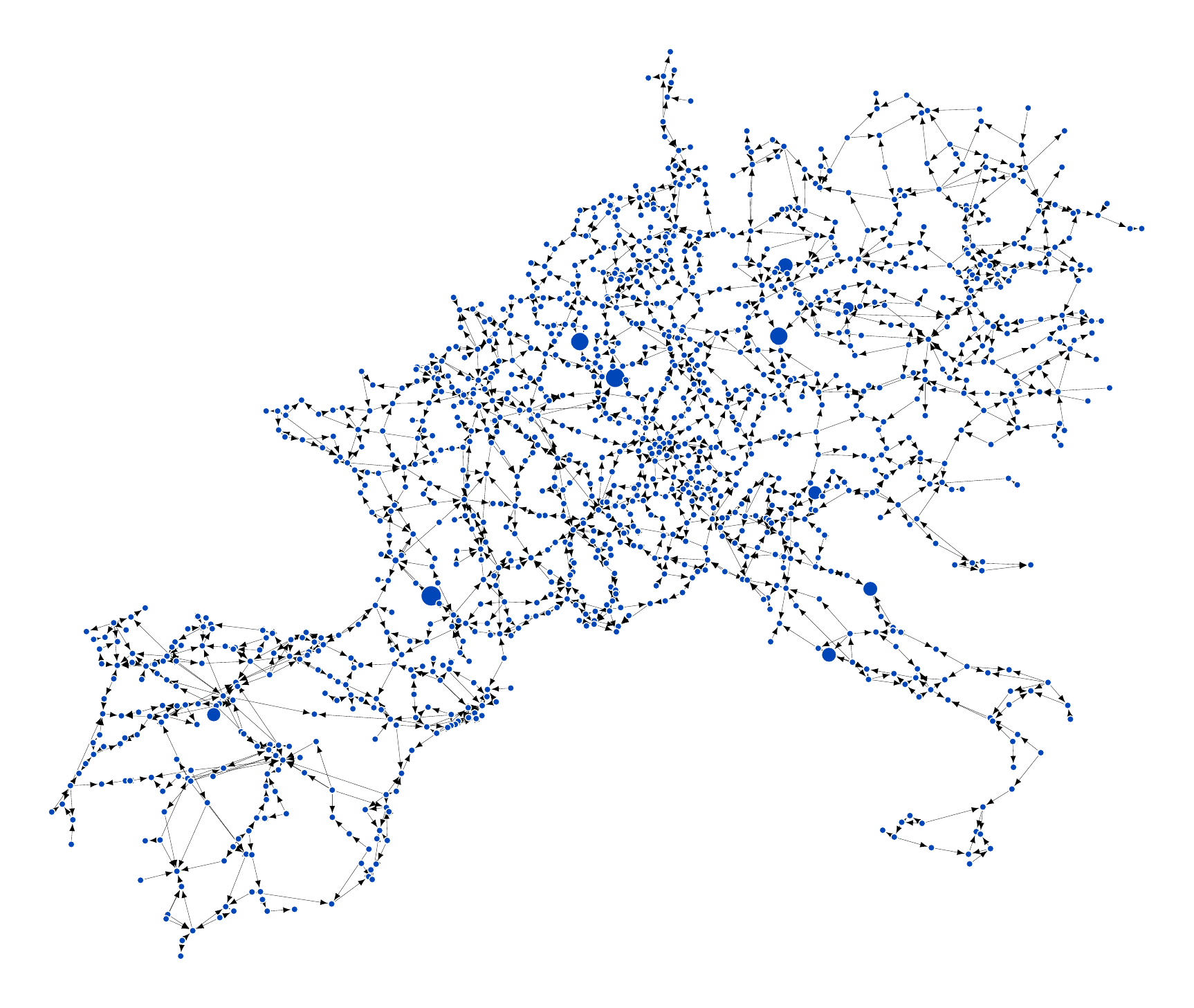}
  \caption{Spatial distribution of load on the European power grid in the steady state for a particular realization of the voltage distribution.
The size of the nodes corresponds to the average load allocated on them.
The smallest size refers to zero load, where the largest one to the maximum load.
}
  \label{fig:power_grid_steady_state}
\end{figure*}

\emph{Perturbation}---Initially, the load on all directed edges and nodes is set to zero.
Perturbations are described as local changes in the load according to the following procedure.
First, we choose a node $i$ at random and set $\rho_i$ to a fixed value $\rho_0$ ($\rho_0>\rho_i$) during a time interval $T_p$.
The load on the corresponding edges and on the other nodes is determined by solving the Burgers equation as described in the section Methods.
After $T_p$, the constraint on the load of node $i$ is released and its load is determined by the dynamics.
A new node is selected and perturbed and the procedure is iterated.
In addition to the perturbations, at each iteration step, 0.1\% of every node load is dissipated.
This dissipation would correspond, for example, to the evaporation of water from a river network, cars leaving the streets, or a potential drop due to Joule heating.

\emph{Directed networks}---The dynamics is investigated on the European high-voltage power grid \cite{zhou05} and two network models: the configuration model with power-law degree distribution \cite{albert02,newman10,bender78,molloy95} and the Watts-Strogatz model, with small-world features \cite{watts98}.
In the case of the model networks, the length of the edges are random variables chosen uniformly from the interval [3:20].
Initially, the power grid and the model networks are undirected.
Inspired by the fact that in power grids the direction of the current depends on the node voltages, we use the following method to define the direction of the link.
To each node $i$, a random value $\phi_i$ (the node voltage) is assigned uniformly from the interval [0:1] and the edge between two nodes is directed from the node with higher voltage to the one with lower voltage (i.e., $\phi_\mathrm{source}>\phi_\mathrm{target}$).
This method for generating directed edges automatically prevents the presence of loops.
Since fluctuations are always present in the network, the direction of the current can vary in time.
Our results are averaged over different voltage distributions as well.

\section{ Results }
\label{sec:results}
\emph{Steady state}\label{subsec:steady_state}---At each time step, we measure the temporal load correlation, defined as:
\begin{equation}
    \Delta\overline{\rho}(t)=\bigg\langle\bigg|\frac{\overline{\rho}(t')-\overline{\rho}(t'-100T_p)}{\overline{\rho}(t')}\bigg|\bigg\rangle_{t'}
    \label{eq:relative_change_in_load}
\end{equation}
where $\overline{\rho}(t)$ is the load averaged over all nodes at time $t$.
The brackets represent an average over the last ten consecutive time intervals of length $100T_p$, i.e., $t'=t-n100T_p$ with $n=0,1,\dots,10$.

Starting with all loads equal to zero, we observe that $\Delta\overline{\rho}$ decays in time towards a steady state in which the dissipation balances the total incoming load.
When $\Delta\overline{\rho}(t)$ drops below 1\% of the relative standard deviation of the loads within the network, we assume that the steady state is reached.
In the steady state, the load at each node has a well-defined average value with small fluctuations.
The spatial distribution for a given realization of voltages in the European power grid is shown in Fig.~\ref{fig:power_grid_steady_state}.
The size of the dots represents the average load over a time window of $1000T_p$ measured in the steady state.
Most nodes accumulate negligible load ($\rho\approx0$), while surprisingly a small fraction of the nodes are overloaded ($\rho>10\rho_0$, where $\rho_0$ is the magnitude of each perturbation).

\begin{figure}[!h]
  \includegraphics[width=8.6cm]{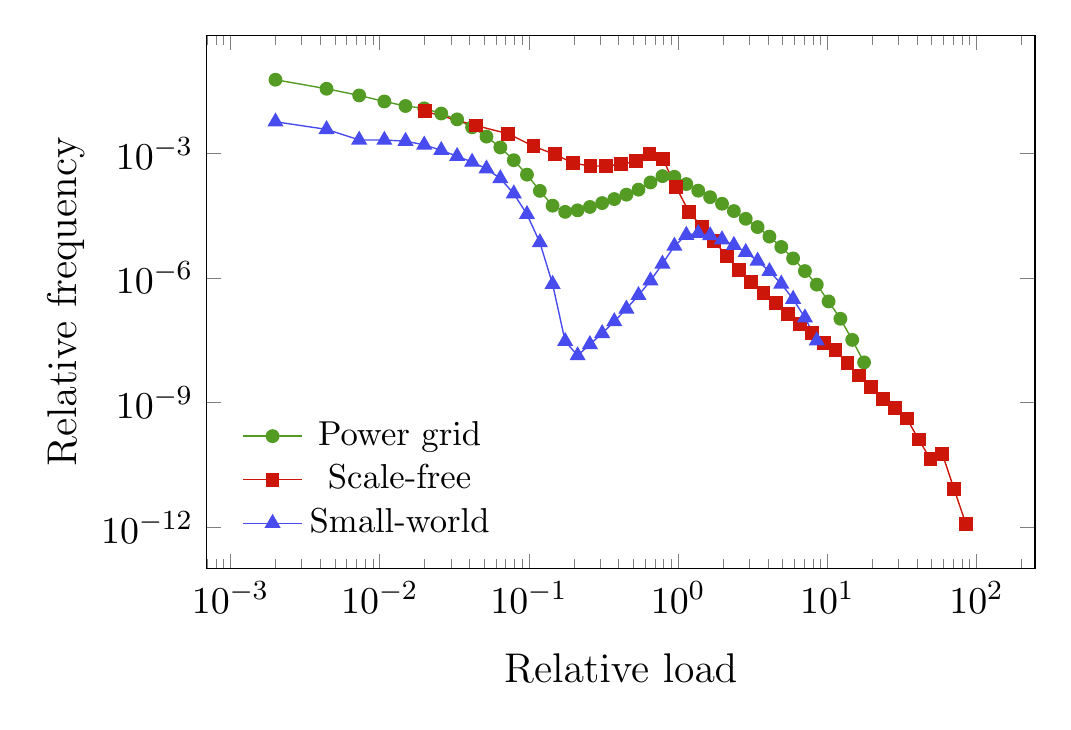}
  \caption{Relative load distribution in the steady state for three network topologies: power grid (green dots), scale-free (red squares) and Watts--Strogatz networks (blue triangles).
The power grid and the scale-free network have $N=1254$ nodes and $M=1811$ edges, while the Watts--Strogatz network has $N=1254$ nodes and an average degree of $\langle k_{out}\rangle=2$.
All loads are in units of the amplitude of the perturbation.
Each curve is an average over 5000 voltage realizations and, in the case of the model networks, also an average over 100 different networks.
The magnitude of the standard deviation of the curves is comparable to the size of the symbols.
}
  \label{fig:steady_state_density_distribution}
\end{figure}

The observed load exemplarily shown in Fig.~\ref{fig:power_grid_steady_state} corresponds to one realization of the voltage distribution and thus for one configuration of the direction of the edges.
Assuming small temporal changes in the network (e.g., fluctuations of voltage in the power grid or number of cars entering a road junction), the direction of the edges changes in time.
Thus, we also consider different realizations of the voltage distribution and, for each realization, we determine the steady state load distribution.
Figure~\ref{fig:steady_state_density_distribution} shows the relative load distribution averaged over $5000$ realizations.
In order to compare the load distribution of different networks (power grid, Watts--Strogatz and scale-free networks), loads in each curve are rescaled by the magnitude of each perturbation ($\rho_0$).
The strongly inhomogeneous behavior of the steady-state load seen in Fig.~\ref{fig:power_grid_steady_state} is also visible in the load distribution.
The distributions are bimodal defining two different types of nodes: those with a negligible load compared to the perturbation ($\rho<\rho_0$) and those with a larger load ($\rho>\rho_0$).
The latter ones are typically overloaded in the steady state, suggesting that the incoming perturbations interfere constructively at them.

\begin{figure}[!h]
  \includegraphics[width=8.6cm]{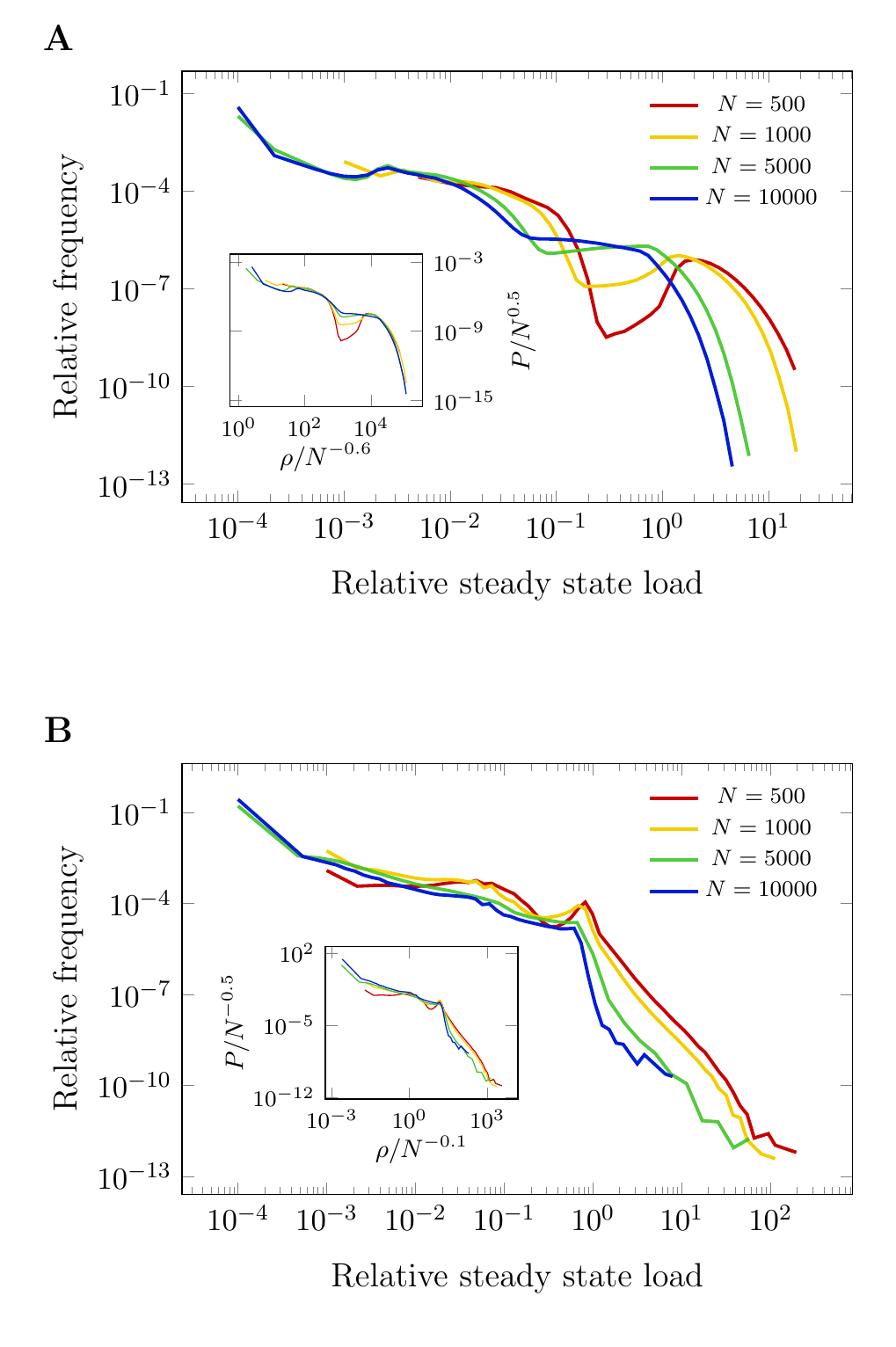}
  \caption{Relative load distribution in the steady state for (A) the Watts--Strogatz and (B) scale-free networks with $\langle k\rangle=2$ for different system sizes, insets show the respective data collapse, where $\rho$ and $P$ denote the load and the relative frequency, respectively, $N$ is the size of the network.
Loads are divided by the magnitude of the applied perturbation to ensure comparability.
Each data is averaged over at least 100 different graphs and 100 voltage realizations.
The breaks in the distributions around $\rho=10^{-3}$ are due to the logarithmic binning.
}
  \label{fig:size_effects}
\end{figure}

The plots for the two network models (Watts--Strogatz and scale-free) in Fig.~\ref{fig:steady_state_density_distribution} are obtained for networks with the same number of nodes as the power grid.
The average degrees are also kept close to the power grid, with the same number of edges in the scale-free network and $\langle k_{out}\rangle=2$ in the Watts--Strogatz graph.
In both cases, a bimodal distribution is also observed.
The power-grid network is constructed from real data and its size corresponds to the real network size.
Thus, a finite-size study is not possible.
Yet, in the case of the model networks one can systematically study the effect of the network size on the load distribution.
Figure~\ref{fig:size_effects}A shows the load distribution for Watts--Strogatz networks of different network sizes.
The majority of the nodes (more than $90\%$) has always a negligible load, while the load of the remaining nodes follows a broad distribution, characterized by a decay in the relative frequency with increasing load and a cut-off for values of load close to $\rho_0$.
The bimodal distribution smoothens out for larger network sizes.
For scale-free networks the qualitative picture is slightly different.
As shown in Fig.~\ref{fig:size_effects}B, for all network sizes, one observes two power-law regimes,
with a crossover at $\rho_0$.
Nevertheless, note that for both network models, there is always a significant fraction of nodes (around $10\%$) with a non-negligible load.
The load value of the cutoff suggests that, at large system sizes, consecutive shock waves that enter the network are separated so that they attenuate their amplitude before being able to interfere.

\begin{figure*}[p!]
  \includegraphics[width=17.2cm]{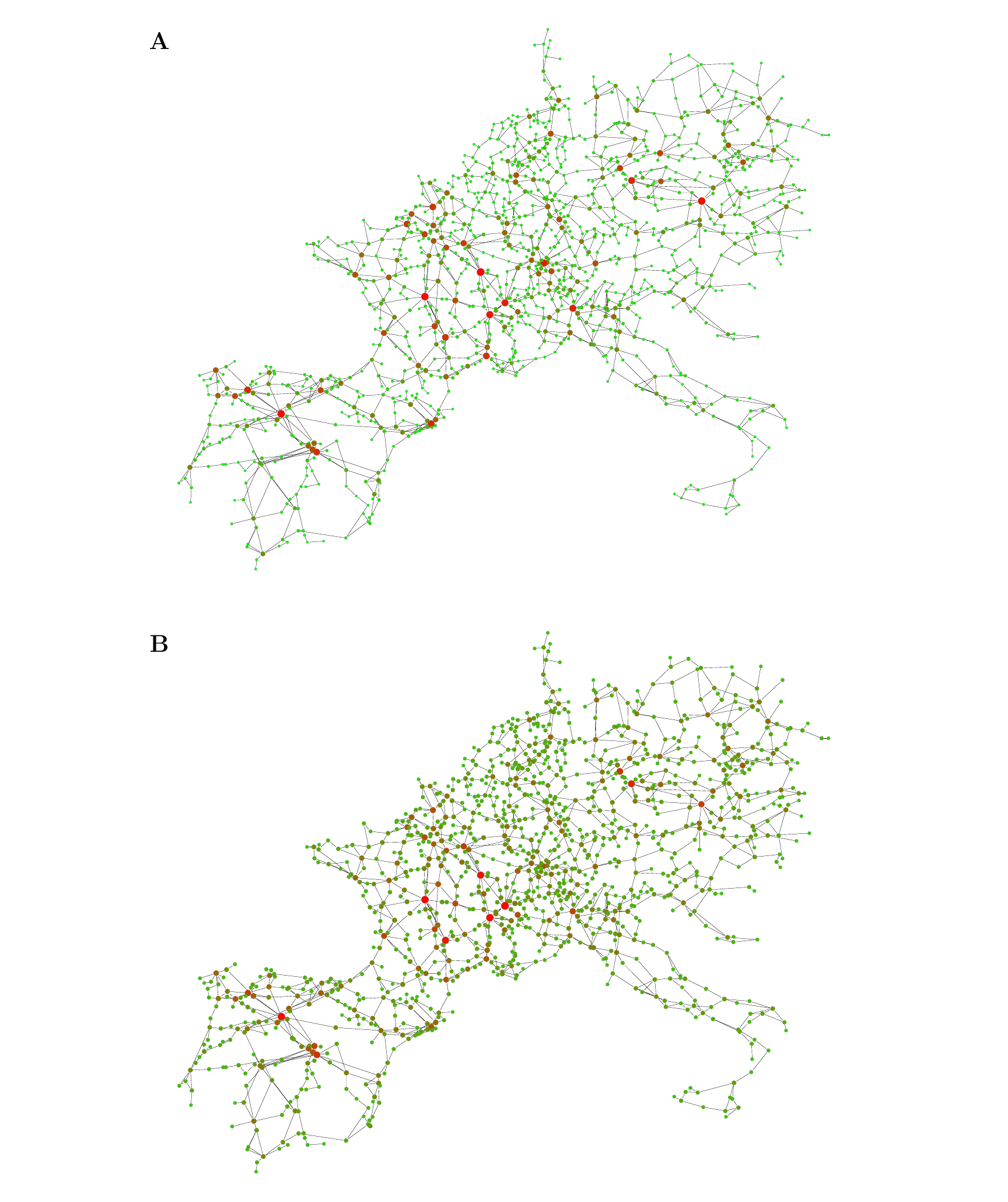}
  \caption{Spatial distribution of two properties on the power grid network.
(A) Vulnerability of the nodes obtained solving the Burgers equation (probability of having $10\overline{\rho}$ in a voltage realization).
(B) Node-basin sizes averaged over different voltage realizations.
Color and size indicate the strength of the corresponding property: red corresponds to large vulnerabilities ($>5\%$) and basin sizes ($>10$), while green nodes have negligible vulnerabilities (below $0.5\%$) and small basin sizes (close to $1$).
The data are the average over 5000 voltage realizations.}
  \label{fig:power_grid_map}
\end{figure*}

\begin{figure}[!h]
  \includegraphics[width=8.6cm]{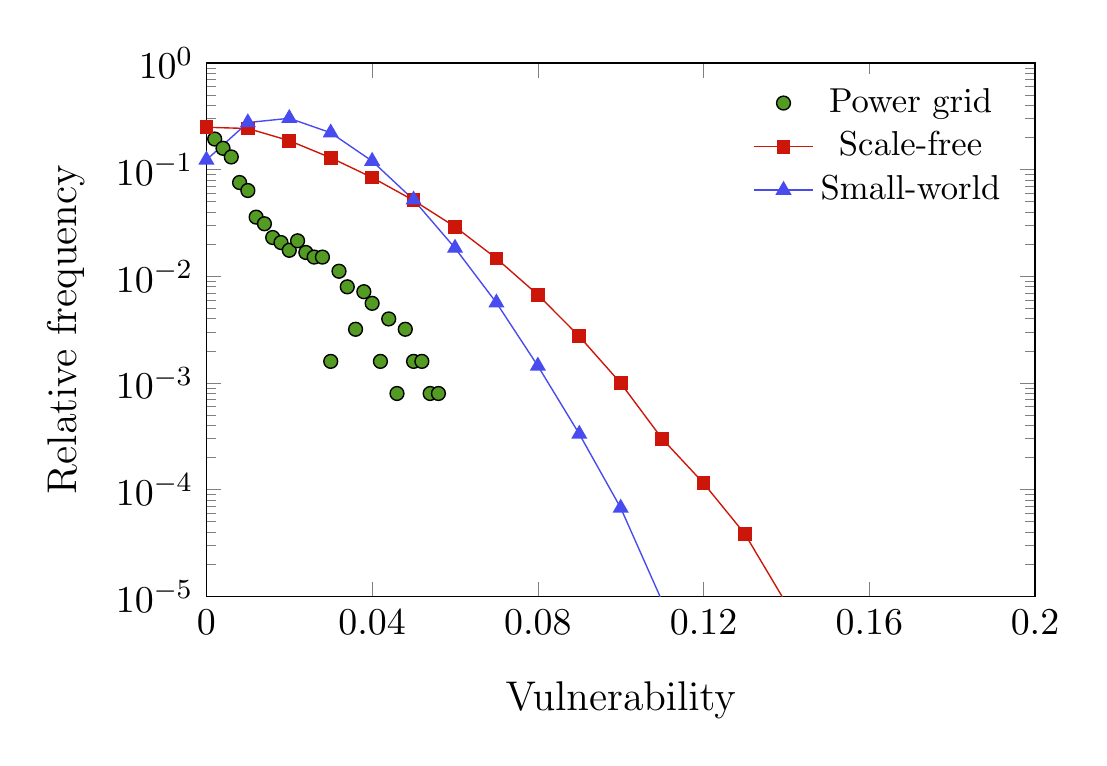}
  \caption{Distribution of the vulnerabilities (i.e., the probability of having a load ten times larger than the average) for three different network topologies: power grid (green circles), scale-free network (red squares) and Watts-Strogatz graph (blue triangles).
The data for the model networks are the average over 100 network realizations.}
  \label{fig:vulnerability_distribution}
\end{figure}

The specific nodes that exhibit these high load values typically change from realization to realization.
However, after averaging over different voltage distributions, we still find some nodes which are consistently overloaded.
For each distribution of voltages, we classify as ``overloaded nodes'' the ones with a load at least ten times larger than the average.
We define \emph{vulnerability} of a node as the probability that it is an overloaded node.
Figure~\ref{fig:power_grid_map}A shows the spatial distribution of vulnerability in the European power grid where the color and size of the nodes denotes their vulnerability.
The vulnerability of green nodes is lower than 0.1\%, while the one of the red nodes is larger than 5\%.
All the other nodes (about 30\% of the nodes, in dark olive color) have a vulnerability between 0.1\% and 5\%.
In comparison, the highly vulnerable nodes are at least 50 times more frequently exposed to large incoming fluxes.
In the case of random perturbations, vulnerable nodes are more likely to fail or be congested.
It is therefore crucial to identify these nodes to improve their capacity and to mitigate the risk of failure.
Figure~\ref{fig:vulnerability_distribution} shows the vulnerability distribution corresponding to the map in Fig.~\ref{fig:power_grid_map}A and to other network topologies.
In the case of the Watts--Strogatz and scale-free networks, a vulnerability distribution for networks of size $N=10^5$ are presented.

\begin{figure*}[!t]
  \includegraphics[width=17.2cm]{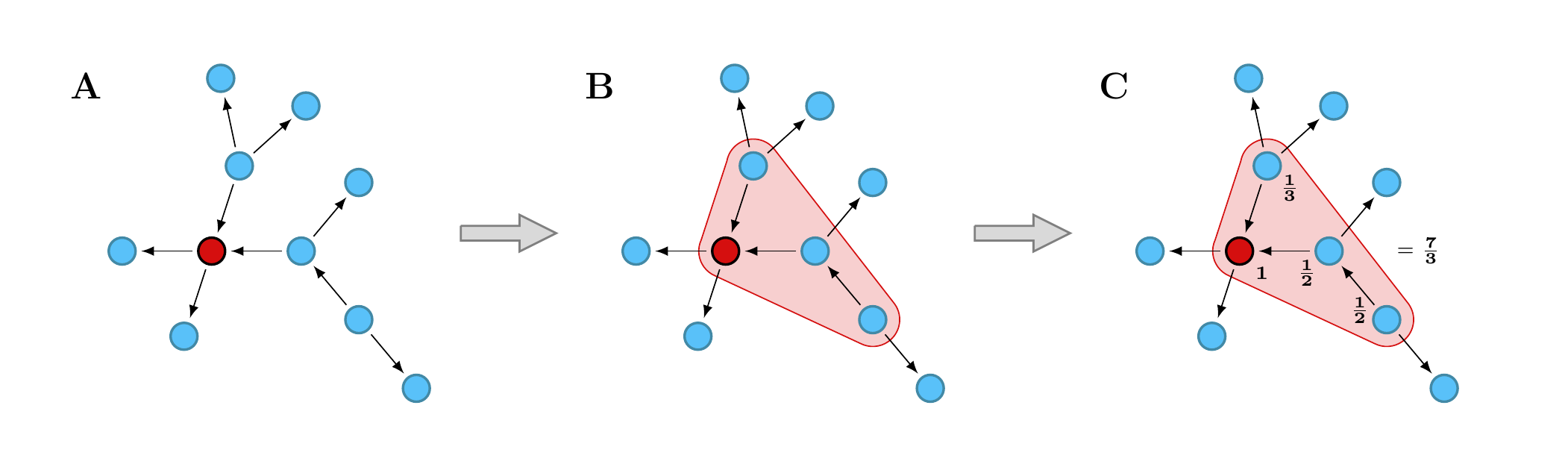}
  \caption{Calculation of the node-basin size.
(A) The basin corresponding to the red node is considered.
(B) We determine the smallest subgraph containing the red node and having only out-edges to the rest of the network.
This is equivalent to a breadth-first search traversing the fraction of the graph reached through only in-edges.
(C) When the basin is determined, each node in the basin contributes to the red node's basin size by the inverse of its out-degree.
The contribution of the red node (i.e., the sink of its basin) is unity.}
  \label{fig:weighted_basin_size}
\end{figure*}

\emph{Identifying vulnerable nodes}\label{subsec:identifying_vulnerable_nodes}---Next we will introduce a simple topological property of the nodes to identify the vulnerable spots without solving the dynamics of the Burgers equation.
According to the Burgers equation, each node sheds the incoming shock wave among its out-edges and the total load agregated at the node at a time $t$ is the sum of incoming loads.
Hence, if we track the path of a given shock wave, it is fragmented at each node with multiple out-edges and will stop at any node that does not have any out-edges.
Assuming that the time average of the load at a node is proportional to the number of incoming shock waves, we determine the \emph{basin} corresponding to that node.
For this purpose, let us consider one realization of the edge directions (see Fig.~\ref{fig:weighted_basin_size}).
For a given node (the one in red in Fig.~\ref{fig:weighted_basin_size}), the corresponding basin is defined as the smallest subgraph of the network containing this node (as the sink), which is connected to the rest of the network by outgoing edges.
Following the procedure as illustrated in Fig~\ref{fig:weighted_basin_size}, we go through the nodes following the opposite direction of the in-edges, starting from the red node (Fig.~\ref{fig:weighted_basin_size}A), and add nodes to the basin that has only out-edges at the end of the process (marked by the blue region in Fig.~\ref{fig:weighted_basin_size}B).
The resulting subgraph is the basin of the red node, and the contribution of the load of the nodes in the basin is simply the inverse of their out-degree (Fig.~\ref{fig:weighted_basin_size}C), except for the initial (red) node, that contributes to the load with unity.
This choice of the contribution of the nodes in the basin is based on the fact that Eq.~(\ref{eq:network_discretization}) conserves the flux at each node.
For simplicity, we assume here that the amplitude of the shock waves leaving a node is on average the same for each out-edge.

\begin{figure*}[!t]
  \includegraphics[width=17.2cm]{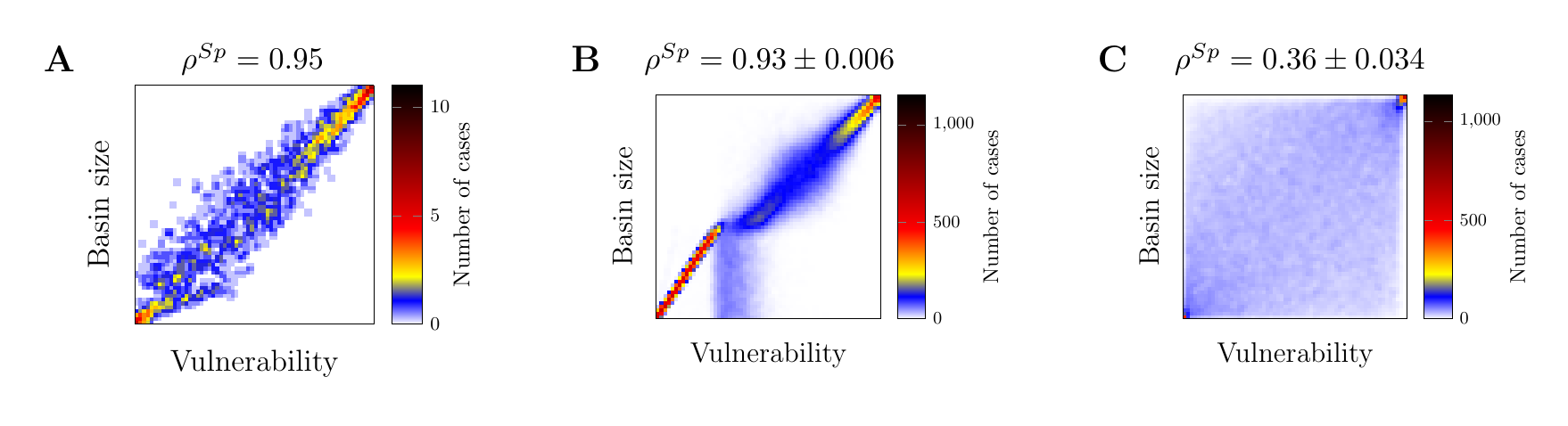}
  \caption{Scatter plots of the ranked vulnerability and node-basin sizes for three different networks: (A) power grid, (B) scale-free and (C) Watts--Strogatz network.
The axes on the plots indicate the corresponding ranked property, and the color denotes the number of nodes for which the two ranks were identical.
In other words, the color of a dot at $(x,y)$ corresponds to the number of nodes with vulnerability rank of $x$ and node-basin size rank of $y$ (see the colorbars).
All plots are the average over 5000 voltage realizations.
In the case of model networks, 100 different realizations are considered.}
  \label{fig:scatter_plot}
\end{figure*}

We calculated the size of the basins for each node, defined as the sum of contributions from all nodes inside the basin of the corresponding node.
This basin size (which is determined for one voltage distribution) is then averaged over different voltage configurations.
The resulting basin size distribution is depicted in Fig.~\ref{fig:power_grid_map}B for the power-grid network.
One sees that for this network, the distribution of the node-basin size is very similar to the distribution of the vulnerability.
A quantitative comparison of the two properties can be given by their correlation.
Thus, we plot the rank-rank scatter plots in Fig.~\ref{fig:scatter_plot} of the indices of the nodes after being sorted in ascending order by vulnerability and node-basin size.
The plots are the average over 5000 voltage distributions and over 100 different topological realizations of model networks.

The corresponding product-moment correlations $\rho^{Sp}$ of the ranks \footnote{The product-moment correlation is, \[\frac{\langle(x-\mu_x)(y-\mu_y)\rangle}{\sigma_x\sigma_y}\] where $\mu_x$ and $\mu_y$ are the mean and $\sigma_x$ and $\sigma_y$ are the corresponding standard deviation of the two quantities $x$ and $y$.} are given above the plots, showing strong correspondence between the ranked vulnerability and node-basin size for the power grid and scale-free networks.
The crucial nodes are the ones with large vulnerability since they are more exposed to large loads.
The basin size shows a strong correlation with the vulnerability for these large values (top right corner of the scatter plots), meaning that it is a good estimator of vulnerability.
Watts--Strogatz network exhibits much less correlation because the degree distribution is extremely narrow, that is, deviations from the average degree are negligible.
Thus, for each realization, the differences in the loads from node to node are very small.
This conclusion is also supported by Fig.~\ref{fig:vulnerability_distribution} showing a narrow vulnerability distribution for Watts--Strogatz networks.

\section{ Discussion }
\label{sec:conclusion}
We study the propagation of shock waves on directed networks using the Burgers equation.
Under sequentially applied perturbations and constant dissipation, the dynamics approaches a steady state.
In this steady state, most of the nodes have negligible average load and a significant fraction of the total load is localized on a few nodes.
We found that some nodes are more likely to accumulate load even after averaging over many edge direction configurations.
These nodes (the vulnerable nodes) are more likely to fail, when there is a finite capacity of the load they can bear.
Unexpectedly we find for the European power grid a broad pronounced bimodal distribution for the loads, while for scale-free network the distribution resembles more a power law.

The steady state and thus the probability distribution of vulnerability among the network is determined by solving numerically the partial differential equations of Eq.~(\ref{eq:1d_burgers_equation}) on each edge.
The propagation velocity of the shock waves depends on their amplitude, which can vary rapidly throughout the network.
We propose a simpler alternative based on the node-basin size to estimate the vulnerability of the nodes and identify the most vulnerable ones.
Simulations on a real network (European high-voltage power grid) and on scale-free networks show that the node-basin size can predict very accurately the location of vulnerable nodes while it performs worse for the Watt-Strogatz network due to its narrow degree distribution.

Our results suggest that it is possible to establish a remarkable connection between dynamics and network structure.
Although for many networks the node-basin size seems to be an accurate tool in predicting the distributions and detecting vulnerability, it is only the first step towards a complete description of the steady state.
More might be understood by studying the relation between the most vulnerable nodes: under what circumstances are they separated or forming connected subgraphs?
Is any local property of the network responsible for a node being highly vulnerable?
This information would provide the tools to mitigate the risk of systemic failure.
Further investigation may involve the removal of nodes that reach their capacity.
In this case, the study of the time evolution of the network structure or optimal strategies of dynamical node/edge addition or deletion can be of relevance.

\section{ Methods }
\label{sec:methods}
\emph{Dynamics}---In this section, we describe the generalization of Eq.~(\ref{eq:1d_burgers_equation}) on a directed network.
The numerical solution of the one-dimensional Burgers equation can be discretized using Godunov's scheme \cite{godunov59,leveque90}
\begin{equation}
    \rho_i^{t+1}=\rho_i^t+\frac{\Delta t}{\Delta x}\big[\overbrace{F(\rho_{i-1}^t,\rho_i^t)}^
                       \mathrm{in-flow}-\overbrace{F(\rho_i^t,\rho_{i+1}^t)}^\mathrm{out-flow}\big],
    \label{eq:numerical_1d_burgers}
\end{equation}
where $\rho_i^t$ is the load at the mesh point $i$ at time $t$, $\Delta x$ and $\Delta t$ are the spatial and temporal discretizations, and $F(\rho,\eta)=\frac{\tilde{\rho}^2}{2}$ is the flux.
The value of $\tilde{\rho}$ is given as follows \cite{godunov59}: If $\rho\ge\eta$ then
\begin{equation}
    \tilde{\rho}= \left\{
    \begin{array}{l l}
        \rho & \quad \text{if $\frac{\rho+\eta}{2}>0$} \\ \notag
        \eta & \quad \text{otherwise} \\
    \end{array} ,\right.
    \label{eq:numerical_flux_1}
\end{equation}
otherwise,
\begin{equation}
    \tilde{\rho}= \left\{
    \begin{array}{l l}
        \rho & \quad \text{if $\rho>0$} \\ \notag
        \eta & \quad \text{if $\eta<0$} \\
        0 & \quad \text{if $\rho\le0\le \eta$} \\
    \end{array} .\right.
    \label{eq:numerical_flux_2}
\end{equation}

\begin{figure}[!h]
  \includegraphics[width=8.6cm]{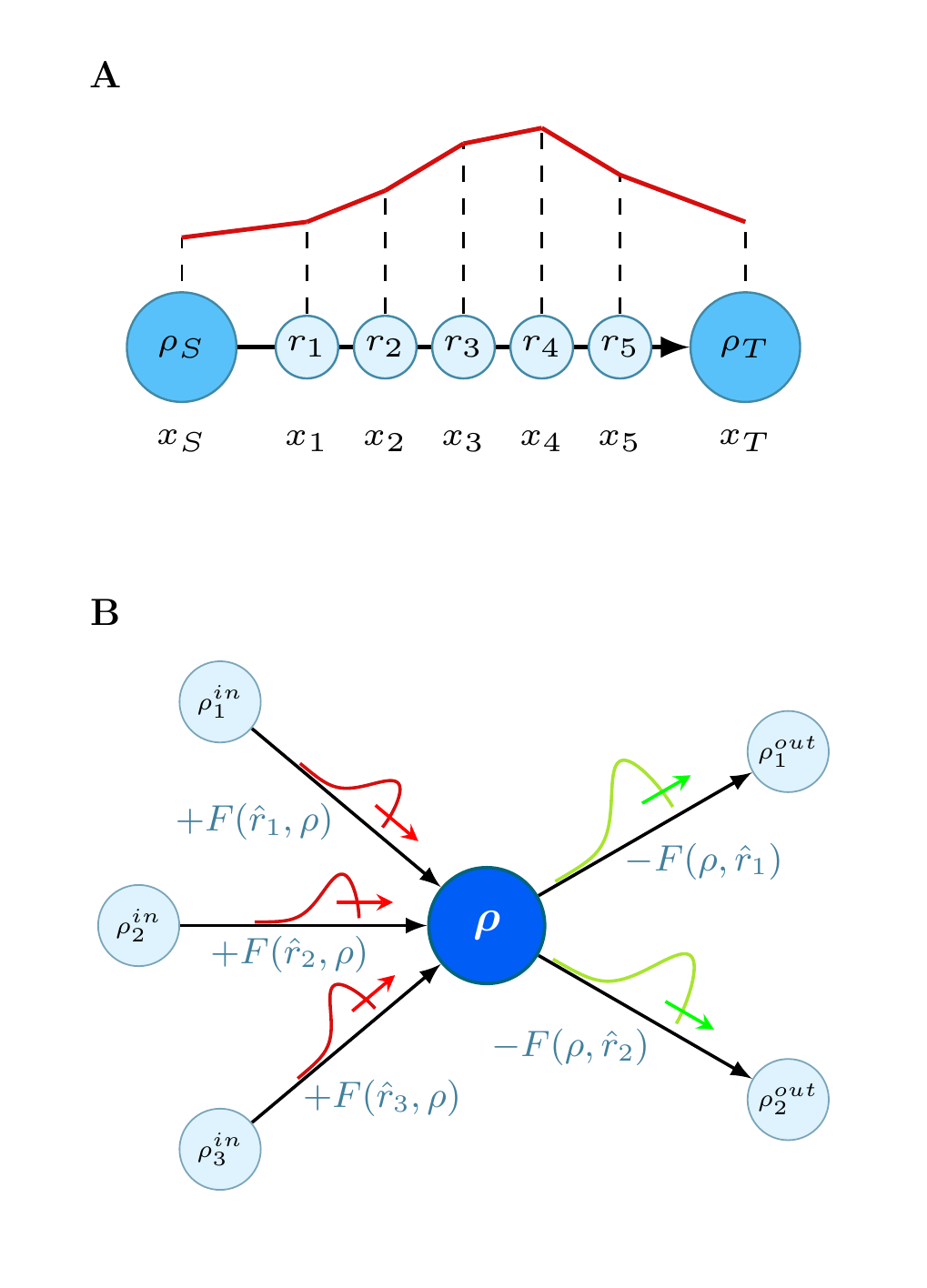}
  \caption{Illustration of the discretization model on a directed network.
           (A) Each edge corresponds to a one-dimensional coordinate system with the positive direction defined by the direction of the edge.
           $x_i$ denotes the $i$th mesh point on the edge, $x_S$ and $x_T$ are the mesh points corresponding to the source and target nodes, respectively.
           The load values on the $i$th edge mesh point are denoted by $r_i$.
           (B) The nodes interact only with the nearest mesh points of their edges.
           $\hat{r}_i$ denotes the adjacent mesh point of the corresponding in- or out-edge of the blue node and $F$ is the flux according to Eq.~(\ref{eq:numerical_1d_burgers}).}
  \label{fig:discretization}
\end{figure}

To solve this equation on a network, one needs to fix the direction of each link in order to have a precise definition of the in- and out-flux of a mesh point.
For practical purposes, this is a realistic approach as water always flows downhill and the current follows a decreasing gradient in electric potential.
Our discretization model for the edges and nodes is illustrated in Fig.~\ref{fig:discretization}.
The edges of a network are one dimensional and thereby Eq.~(\ref{eq:numerical_1d_burgers}) holds.
The number of mesh points in each edge is proportional to its length and the direction is defined by the direction of the edge \footnote{In the case of model networks, the length of the links is uniformly distributed random values}.
Furthermore, a value of $\rho_k$ is assigned to each node $k$.
Nodes interact with the nearest mesh points of their incident edges according to the following equation,
\begin{eqnarray}
    \rho_i^{t+1}\,=\,&\,\rho_i^t\,&\,+\,\overbrace{\frac{\Delta t}{\Delta x}\sum_{j\in B_i^{in}}F(\hat{r}_j^t,\rho_i^t)}^\mathrm{in-edges} \notag \\
                     &            &\,-\,\underbrace{\frac{\Delta t}{\Delta x}\sum_{k\in B_i^{out}}F(\rho_i^t,\hat{r}_k^t)}_\mathrm{out-edges}.
    \label{eq:network_discretization}
\end{eqnarray}
$B_i^{in}$ ($B_i^{out}$) denotes the set of in- (out-) edges of node $i$, and $\hat{r}_j$ ($\hat{r}_k$) is the load at the last (first) mesh point of the corresponding edge. The resulting dynamics conserves the total mass and at each node, the total incoming flux is equal to the outgoing one.

Also, we should add an important remark on the various constraints of the investigated model.
First, in the numerical solution of a nonlinear PDE on a network, the degree of a node corresponds to the local dimension of the space in which one solves the equations.
As the size of the scale-free network increases, the frequency of nodes with very large degrees also grow.
Considering numerical stability, the appearance of larger degrees sets an upper limit on the magnitude of the applied perturbations.
However, if the perturbations are small (which is required by the numerical treatment), shock waves tend to vanish by travelling on the edges and they are not able to interfere constructively.
Therefore, in the finite-size study, we considered only networks below the size of $N=10^4$.

\emph{Networks}---We consider three network models: the European power grid (with $N=1254$ nodes and $M=1811$ edges, i.e., $\langle k_{out}\rangle=1.44$), the Watts--Strogatz small-world network ($\langle k_{out}\rangle=2$) and scale-free network models ($\langle k_{out}\rangle=1.44$ for $N=1254$ and $\langle k_{out}\rangle=2$ in the case of large network sizes).
The Watts--Strogatz network is constructed by considering first a one-dimensional chain with first and second neighborhood connections and periodic boundary conditions, and then rewiring each edge with probability $p=0.01$ (with undirected edges, this corresponds to an undirected average degree $\langle k\rangle=4$).
After the voltages are set and edge directions are introduced, the resulting network has a directed average degree of $\langle k_{out}\rangle=2$.
The scale-free network is constructed by the configuration model: first we assign the degrees for each node according to a power-law with exponent $\gamma=2.5$, and then connect randomly chosen nodes.
Finally, further rewiring of the edges is carried out in order to eliminate degree-correlations.
Note that the number of edges in the Watts--Strogatz network is different from that in the power grid and the scale-free network.

\section{ Acknowledgments }
\label{sec:acknowledgements}
Authors would like to thank the Swiss National Science Foundation under contract 200021 126853, the ETH Z\"{u}rich Risk Center for financial support.
This work was also supported by the European Research Council (ERC) Advanced Grant 319968-FlowCCS and the EU ERC FP7 COLLMOT Grant No: 227878.

\providecommand{\noopsort}[1]{}\providecommand{\singleletter}[1]{#1}%

\end{document}